# CIM compliant multiplatform approach for cyber-physical energy system assessment

Minh Tri LE[1], Van Hoa NGUYEN[1], Quoc Tuan TRAN[2*], Yvon BESANGER[1*],
Thierry BRACONNIER[1], Antoine LABONNE[1], Hervé BUTTIN[2]

[1] Univ. Grenoble Alpes, CNRS, Grenoble INP, G2Elab, F-38000 Grenoble, France
[2] CEA-INES, 50 Avenue du Lac Léman, 73370 Le Bourget-du-lac, France
[*] Senior Member, IEEE
Email : van-hoa.nguyen@g2elab.grenoble-inp.fr

*Abstract—* **With high penetration of distributed renewable energy resources along with sophisticated automation and information technology, cyber-physical energy systems (CPES, i.e. Smart Grids here) requires a holistic approach to evaluate the integration at a system level, addressing all relevant domains. Hybrid cloud SCADA (Supervisory, Control And Data Acquisition), allowing laboratories to be linked in a consistent infrastructure can provide the support for such multi-platform experiments. This paper presents the procedure to implement a CIM (Common Information Model) compliant hybrid cloud SCADA, with database and client adaptive to change in system topology, as well as CIM library update. This innovative way ensures interoperability among the partner platforms and provides support to multi-platform holistic approach for CPES assessment.**

*Index Terms—* **Cyber-Physical Energy System, Interoperability, CIM, Cloud SCADA, Holistic Approach.**

## I. Introduction

The European low-carbon energy policy imposes a mandatory requirement on the integration of a high number of distributed Renewable Energy Sources (RES) into the conventional power network [1]. The utilization of these energy sources creates considerable challenges to power system operator in ensuring the power supply dependability and power quality due to their intermittent attributes and low storage capabilities. Moreover, the interaction of the electronically interfaced RES with the grid utilities through dedicated communication channels causes the future so-called smart grids even more complicated. The dependable and optimal operation of such cyber-physical energy systems (CPES) requires advanced automation and information technology as well as corresponding control strategies and data analysis techniques [2], [3].

A holistic research and development methodology for such CPES must not only handle the entire development process (design, analysis, simulation, experimentation, testing and implementation) but also consider all relevant components; aspects and impacts involved in the future power system that might perform interactive activities with controlling devices, algorithm(s), or application case in question. In fact, the results of testing a system with high level of integration of intermittent generation sources are regarded invalid if potential disturbances caused by customers, markets; ICT availability, etc… are not taken into consideration. However, it is not feasible to conduct extensive analyses of these enormously sophisticated and highly integrated systems [3]. Therefore, it is required to develop rigorous testing strategies so as to support the confirmation of integrated systems of various domains represented at different research infrastructures (RI).

In spite of the theoretical possibility of a functional integration of the above-mentioned RI that operates simultaneously and creates integrated holistic energy system, it remains virtually impossible. In order to possess capability to carry out tests and experiments typical of integrated smart grids, testing and experimentation must be possible across distributed and not essentially functionally interconnected RI. To reach that objective, a top priority should be given to the interoperability among their infrastructures.

Thanks to the interoperability among RI, researchers at different areas are able to gain access to shared resources and utilize technical facilities of the partners for research purposes. Interoperability among RI is extremely important for applications that require coordination of facilities from respective platforms, e.g. hardware or power-hardware-in-the-loop co-simulation. Joining existing interoperable infrastructures is much more time-saving and requires much fewer resources compared with building new essential testing platforms. The common realization and control of technological devices offer also opportunities to achieve multi-site research projects, e.g. wide area monitoring and protection, large-scale power management, etc. The ability of joint operation of various platforms enables users not only to

This research is sponsored by the French Carnot Institute "Energies du Futur" (http://www.energiesdufutur.eu/), in the PPInterop 2 project.
The participation from G2Elab and CEA INES is also partly supported by the European Community's Horizon 2020 Program (H2020/2014-2020) under project "ERIGrid".

exchange and deploy useful data amongst power networks and automation systems but also to present and control in real-time the available experiment tools at the partner sides. It produces a significant effect on the expense of constructing and integrating experimental modules at different areas. Additionally, interoperability of RI also facilitates the connection and integration of new platforms [4].

In [5], a novel hybrid cloud SCADA architecture with Software-as-a-service (SaaS) and Platform-as-a-service (PaaS) deliverance models was proposed to link multiple smart grid platforms in a consistent manner, providing the support for a holistic approach of cyber-physical energy system assessment, while assuring reliability and security of electrical services as well as confidentiality for information exchange. Recent information models, using object oriented approaches, such as IEC 61850 [6], Common Information Model (CIM – IEC 61970/ IEC 61968) [7] and OLE for Process Control (OPC UA - IEC 612541) [8] abstract data model are employed in the proposed architecture [5] to ensure a seamless and secured interconnection among platforms.

Whereas the architecture promises a wide range of multi-platform applications toward a holistic approach for CPES assessment, the juxtaposition of various standards in different interoperability layers requires a considerable effort from the system manager. While there exists tools for conversion among IEC61850, CIM and OPC UA, such as CIMBat or UMLBaT [9], creating CIM compliant database as well as online configuration of system topology remains a tedious task, especially when it comes to keeping the CIM version up-to-date, inside platform and for communication with partners. In this paper, we contribute to the architecture proposed in [5] with algorithms allowing the extraction of system topology information from CIM/XML (eXtensible Markup Language )/RDF(Resource Description Framework) schema and auto-creation of database tables based on the extracted information. This contribution provides a CIM compliant hybrid cloud SCADA implementation that enables multi-platform CPES assessment.

In the following, section II reviews the necessity of a multi-platform approach for CPES assessment, highlights the technical challenges of such an approach and why the hybrid cloud SCADA architecture is suitable. Section III recalls the proposed architecture and presents the algorithms for processing CIM information models as well as data assessments. Conclusions and outlooks for further development are drawn in last section.

II. MULTI-PLATFORM APPROACH FOR CPES ASSESSMENT

Due to the integration of Distributed Energy Resources (DER), the conventional distribution network has become a bidirectional power-flow grid, resulting in lots of disadvantages for the existing communication infrastructure such as sectionalized network topology, data overload due to the lack of sufficient bandwidth for two-way communication as well as impossibility to process huge amount of data from smart devices like smart meters, Intelligent Electronic Devices (IED) [10]. Complex CPES like smart grids are operated based on a combination of different advanced technologies in various areas (power, information and communication technologies (ICT)/automation, markets, customer response, etc.) that create mutual interaction and interdependencies. Prior to implementing algorithms and solutions, field studies should be conducted in order to assess the interoperability among systems, handling all domains involved. Hence, the careful consideration of communication network is of vital importance to development of smart grids with regard to efficient topology, impact of communication latency, and security.

The methodologies and technologies developed for smart grids have created an increasing need for a more interactive simulation approach involving all integrated parts[10]. A general framework for smart grid validation and launch is of necessary. However, this is faced with several obstacles, one of these is the shortage of design methods and related software tools capable of simulating power and ICT networks simultaneously [3].

The communication latency has very limited impact on the results of experiments conducted in the lab because of rather short-distance communication channels. As a result, this does not depict the real situation where a long-distance communication channel may produce undesired delay and signal loss and may cause mal-operation for timely control. Thus, in order to study the effects of communication platforms on the dependability and performance of the supervisory and control systems as well as the scale of the ICT infrastructure needed to invest to meet the requirements, the communication network are often carried out separately using dedicated simulator software[11]. Also, these communication simulators simplify cyber-security related studies such as denial-of-service protection, confidentiality and integrity checking, that are rather important but not always simple for the power and energy community.

Up-to-now the development of such a cyber-physical approach for designing, analyzing, and validating smart grid systems is still gated by the laboratory's limited infrastructure and the domain-bounded capacity of staff. The existing laboratory based testing approaches often focus on a certain sub-system (or business sector) and its components. Linking laboratories with complementary expertise and infrastructures, in a consistent platform, will allow conducting and analyzing CPES experiment in a holistic manner. It is however, not an easy task, mainly due to the necessity of synchronizing both platforms properly at runtime. Moreover, the existing devices and simulation tools offer limited options of adequate Application Programming Interface (API) for external coupling. On top of that, the fundamentally different concept behind power system and communication networks is also a challenge to detect, link, and handle related events in both sides [12]. In the case of CPES assessment, the requirement is even higher due to the nature of some applications (e.g. protection) that require strictly fast and deterministic response time. Moreover, when it concerns multi-laboratory aspect, cyber-security assessment for information exchange also needs to be considered.

The hybrid cloud architecture proposed in [5] offers solutions for the aforementioned challenges. In this method, the local SCADA server, Programmable Logic Controller (PLC) or Remote Terminal Unit (RTU), etc. is used to control the platform control network, to which the SCADA application that runs on-site is directly connected. This maintains an appropriate delay (as the communication is conducted via Local Area Network - LAN) and high level of security (the data exchanged locally) for the crucial functions. A common committee and technical personnel responsible for the interoperability with the network should strictly moderate the (physical or virtual) common cloud SCADA server. Adopting Platform-as-a-Service (PaaS) transfer model, distant permitted users, collaborative institutions, and a number of applications can obtain the visualization, report and limited access to the particular selected non-critical information which is transferred to the common cloud SCADA service (Figure 1). As stated in platform-linking agreement, a remote user that is authorized by the possessor of the platform can make a request to run a certain task (setting thresholds, executing simulation, etc.) and visualize the results. This functionality is very crucial in the context of interoperability amongst RI platforms as it provides the opportunity for researcher at one place to conduct tests using the remote shared resources.

The architecture enables the possibility to cooperate assets and expertise from various RI in multi-platform experiments for CPES assessment. In the following section, we present our adjustments to the architecture to make it CIM compliant and to create a self-adapted database with respect to CIM version update.

### III. CIM COMPLIANT HYBRID CLOUD SCADA IMPLEMENTATION

#### A. Procedure

One of the challenges of the multi-platform hybrid cloud SCADA approach is to enable auto-configuration of the topology of the considered CPES on-the-go and to update the database accordingly, with the information in the CIM/XML/RDF file received from the local SCADA. Moreover, CIM is an ongoing standard with regular updates. Keeping the information model up-to-date is important to ensure the interoperability among partners of the network, as well as in the case of addition of a new partner. A procedure (Figure 1) of implementation of the hybrid cloud SCADA server and database with user interface adaptive to system's topology is then proposed to ease the aforementioned difficulties.

The main difference of this proposal with the classical way of implementation of CIM standard is that the information model is not hard-coded into the database and the hybrid cloud SCADA, but is created and adapted on-the-go via the RDF schema received from the local SCADA server, with respect to the stored CIM library.

The procedure (Figure 1) is composed of 6 tasks:

1. Getting the CPES configuration from the local SCADA server via a CIM/XML/RDF file.
2. Extract the RDF schema for information on system elements and interconnections.
3. Scanning the CIM library for attributes associated to the required elements.
4. Create tables in the database for the required elements, according to the gotten attributes and the interconnection information. In current implementation, SQL (Structured Query Language) server is chosen out of personal preference. However, the algorithm should be easily adaptable to other database management tools. The User Interface (UI) on client side is updated accordingly to any changes in the cloud SCADA.
5. Map the ID from the hybrid cloud database to the ID provided by the local server SCADA inputs for dynamic exchange.
6. Actualize the dynamic connections among the SCADA servers with adequate refresh rate.

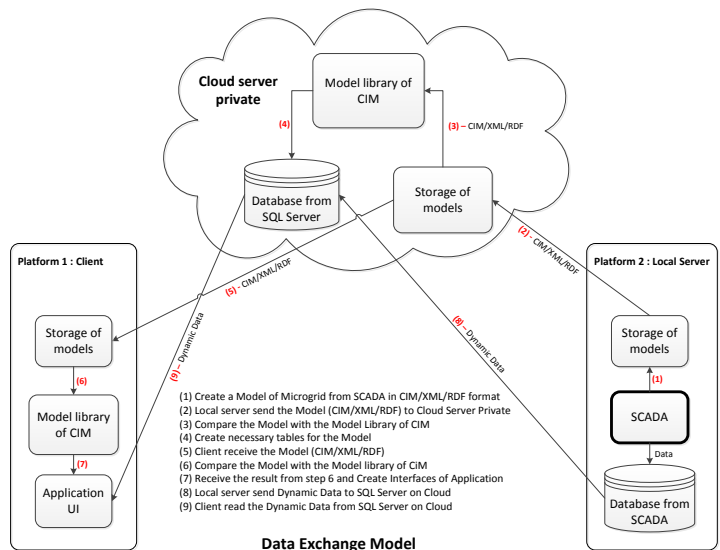

Figure 1: The proposed procedure of implementation of CIM compliant hybrid cloud SCADA system for multi-platform CPES assessment.

The proposed procedure minimizes the effort needed to keep the information model up-to-date (thanks to the independently located CIM library), as well as ensures that the database is adapted to any changes in system topology, which is one of the most difficult issue for CPES assessment in general.

The following sections provides the core algorithms for the above tasks: auto-extraction of information on system elements and their interconnection, creation of the SQL database according to the extracted information and the CIM library and reconfiguration of client UI to system topology.

## B. CIM/XML/RDF auto-extraction

In order to create or adapt the database, the hybrid cloud SCADA server makes a request of CIM/XML/RDF system configuration to the local SCADA server (periodically or per change in topology). Once receiving the feedback, it proceeds to extract the RDF schema for the connected devices (Figure 2). The system then performs a check in the CIM library (Figure 3) for associated attributes and functions. In the current implementation, the library is stored in form of .xmi file, provided by CIM User Group[1]. Utilizing the algorithm for different libraries requires several minor tweaks (i.e. file name, tag name, child level, etc.).

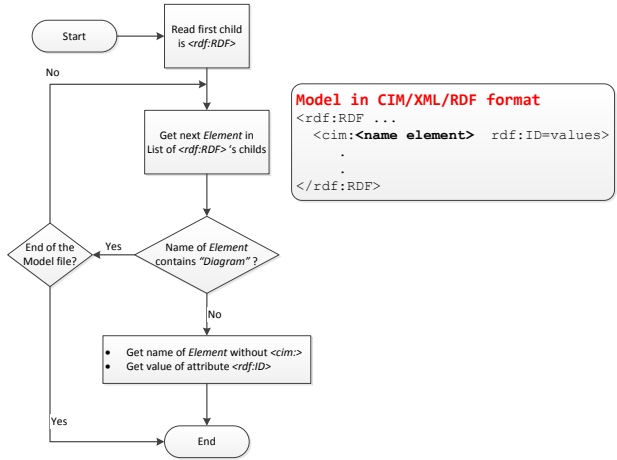

Figure 2: RDF schema processing algorithm

Due to the way the CIM library is stored, one may need to get the attributes and the according types separately (Figure 4 and Figure 5). The information is then compared with the actual database to detect any changes in system topology and missing or redundant attributes. This scan helps to keep the consistency among all the platforms in the network, as well as keep the system up-to-date.

The Client UI is then reconfigured to adapt to the changes.

## C. Auto-creation of database based on CIM library

The database is auto-created by performing a check on *mRID* and *<rdf:ID>* from the received RDF schema (Figure 6).

Using a unique ID to configure the database, we can avoid potential conflicts in the database when the topology of the system changes, or when the SCADA operator decides to update the CIM version. In the latter case, however, a re-initialization of the application is necessary, since data structure may drastically changes.

## D. OPC UA- SQL interface

In [5], OPC UA was chosen as the protocol for SCADA system. The integration of information from the OPC UA server (local SCADA server) to the SQL server (the hybrid cloud server) is important to ensure a seamless dynamic exchange. This is however not in the scope of this paper because several open-source and commercial solutions for such integration already exist.

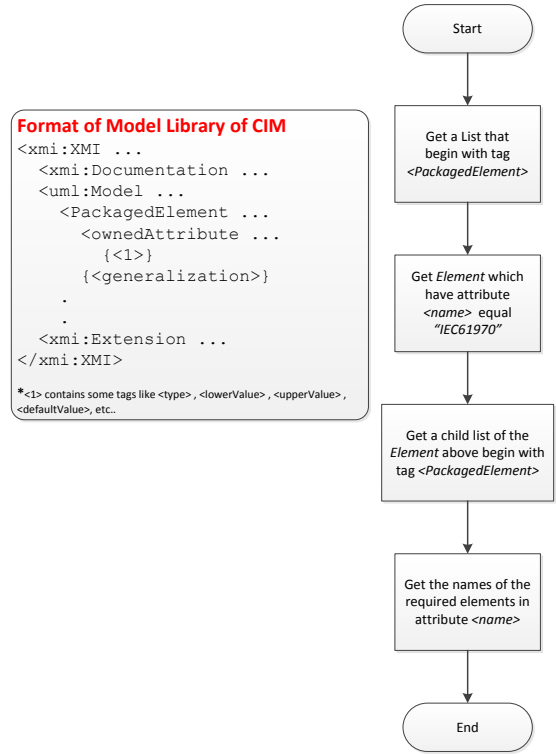

Figure 3: CIM library assessment

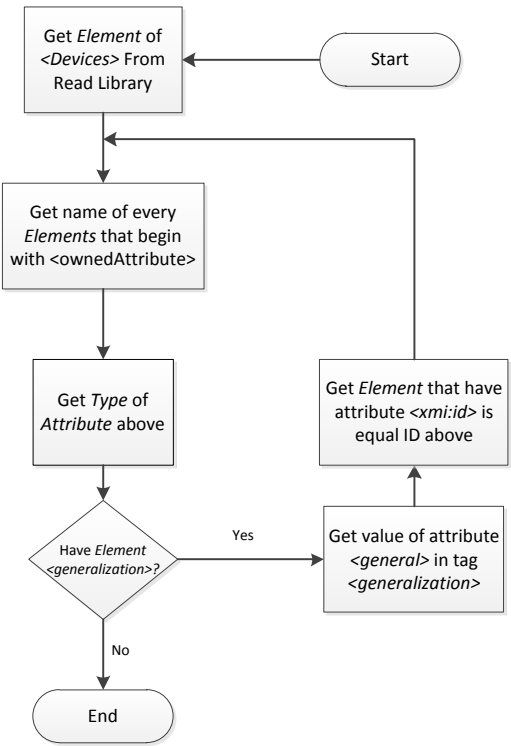

Figure 4: Get attributes of a device from CIM library.

---
[1] http://cimug.ucaiug.org

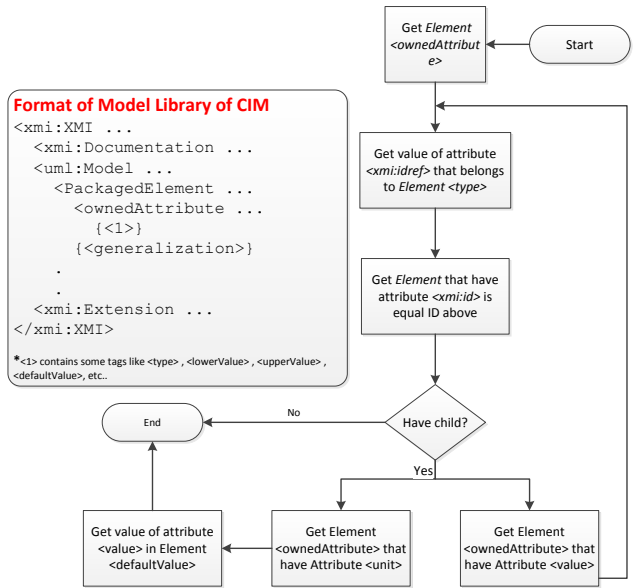

Figure 5: Get Type of an attribute in CIM library.

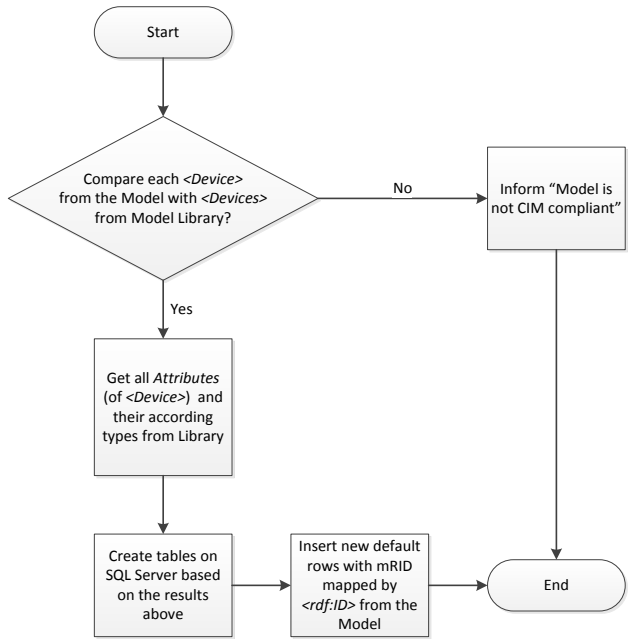

Figure 6: Auto creation of SQL database based on the system configuration extracted from RDF schema and compared to CIM library.

### E. UI auto-reconfiguration

The user interface is designed to be auto-reconfigurable accordingly to system change as well as CIM library updates (Figure 7). In the first version, all the devices are represented as a button leading to their proper online data sheet. A graphical representation is in development.

### F. Implementation example

A demonstration of the procedure on a very simple system is represented in Figure 8, with the illustration of corresponding modules. The UI of the client, adapted to the considered test-case is introduced in Figure 9.

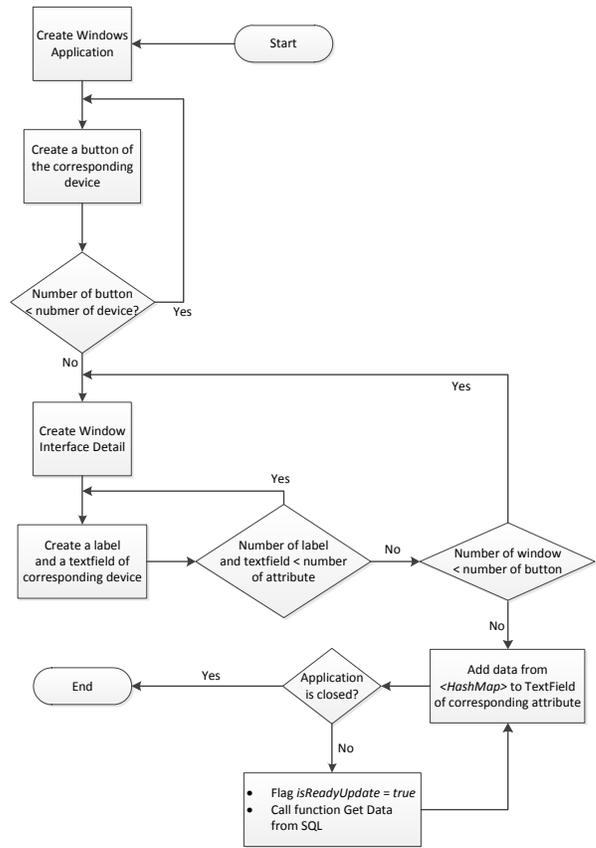

Figure 7: Algorithm for the adaptive UI.

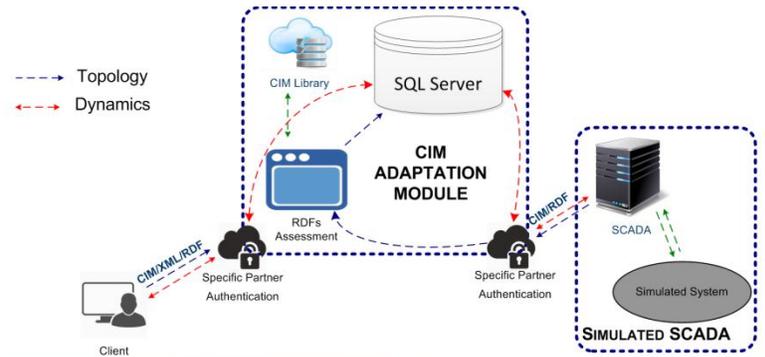

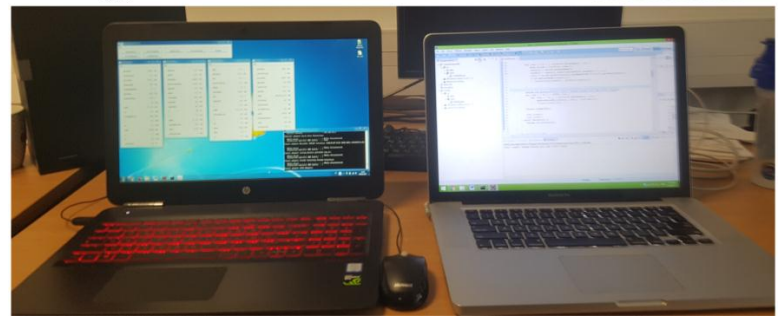

Figure 8: Implementation of a CIM compliant adaptive SCADA for a simple system, with corresponding modules.

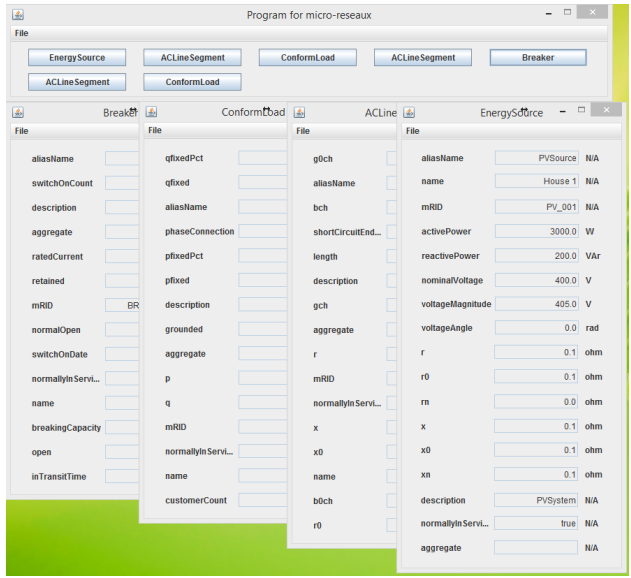

Figure 9: Adaptive UI of the client in the considered test-case.

IV. CONCLUSION AND OUTLOOK

Not only at physical layer, but also interoperability among partner's infrastructure in information model layer is an important prerequisite condition. The Common Information Model (CIM) – IEC 61970 and 61968 standard – is capable to assure such interoperability. However, while there is a lot of effort to implement CIM in academic and industrial scale, migrating the existing information models to CIM is still not an easy task, mostly due to the over-complexity of CIM and due to its ongoing evolvement. In this paper, a procedure to implement a CIM compliant hybrid cloud SCADA was presented as a support to perform multi-platform CPES assessment experiment. The database on server side and the UI on client side are adaptive to the change of system topology, informed via a RDF schema from the local SCADA server, as well as to any update of the CIM library. This way of CIM implementation facilitates drastically the task of system operators.

This innovative way ensures interoperability between partner platforms and provides support for a holistic multiplatform approach to smart grid evaluation. This is a very important contribution to the refinement and improvement of the proposed interoperability architecture via hybrid cloud SCADA. Flexibly assuring interoperability of the laboratory linked infrastructure, various innovative ways of exploiting this new system for CPES assessment can be considered: multi-platform co-simulation or remote Power-Hardware-in-the-loop.